# Liquid crystalline phases and demixing in binary mixtures of shape-anisometric colloids

*S. D. Peroukidis, A. G. Vanakaras and D. J. Photinos*

Department of Materials Science, University of Patras, 26504 Patras, GREECE

# ABSTRACT

A theoretical model of shape-anisometric particles embedded in a cubic lattice is formulated for binary mixtures combining rod-like, plate-like and spherical particles. The model aims at providing a tool for the prediction and interpretation of complex phase behavior in a variety of liquid crystalline colloids, biological and macromolecular systems. Introducing just repulsive interactions among the particles, a rich variety of phase structures and multiphasic equilibria is obtained, including isotropic, nematic, lamellar and columnar phases, demixing into phases of the same or different symmetries and structural microsegregation of the different species of the mixture within the same phase.

# 1. Introduction

Understanding the structure-properties relations of complex molecular systems is a central concern in the field of soft condensed mater and biology. The physicochemical properties, the response in external stimuli and/or the biological function of a wide class of soft molecular systems_ including thermotopic and lyotropic liquid crystals, dendrimers, colloidal dispersions, block copolymers and bio-macromolecular systems_ depend critically on their intrinsic ability to spontaneously self-organize into macroscopically ordered mesophases. Understanding the underlying molecular mechanisms and interactions of spontaneous self-organization is a central scientific and technological challenge towards controlling and optimising the functions of existing materials and a key step in the molecular engineering of new materials.

Colloidal suspensions of spherical or anisometric particles and their mixtures offer a vast variety of order-disorder phase transitions. The well defined particle shapes and the possibility to control the interparticle interactions render colloidal suspensions the mesoscale analog systems for the study of phase transitions in molecular systems. A classical example is provided by the seminal work of L. Onsager on the liquid crystal behavior of anisotropic rod-like colloids[1] and its subsequent great impact on the development of molecular theories and models for computer simulations of low molar mass thermotropic liquid crystals.

There are several experimental studies on the spontaneous self organization and on the phase decomposition of mixtures of rod-like and spherical particles[2-6], where the spherical particle may be colloidal spheres[2-6] or biomacromolecules[3] or nonadsorbing polymers[2-6]. Apart from the isotropic phase, these mixtures may exhibit uniform nematic phases and lamellar phases or, depending on the relative concentration of the two species, they may demix into phases of different symmetry. For example, a broad isotropic-nematic coexistence region has been observed[7] at low temperatures in a temperature-concentration phase diagram of a mixture of rods (low molecular mass LC) and larger particles (silicon oil). Also, Lekkerkerker et. al. investigated mixtures of colloidal silica spheres with smaller colloidal silica rods and found that lowering the rod concentration could facilitates sphere crystallization[8-10].

Phase structures and transitions in mixtures of plate-like colloids with spherical particles have been explored experimentally[11-13]. Self-organization into uniform columnar phases, even for relatively high sphere concentrations, as well as the possibility of demixing into phases of the same symmetry is found in such mixtures. There are also experimental studies on the phase behavior of mixtures of highly asymmetrical rod-like and plate-like colloids[14,15]. These systems exhibit a rich phase polymorphism which covers nematic, smectic and columnar order as well as separation into coexisting phases of any of the above symmetries. A particular interest for the nematic regime of rod-plate mixtures has been stimulated by the possibility of formation of a biaxial nematic phase[16,17].

The experimental exploration of mesomorphic mixtures of anisometric particles has motivated extensive theoretical and computer simulation studies[17-49]. The fragile nature of these states of soft matter is a result of relatively weak intermolecular interactions[50] which, despite their weakness, are able to support the macroscopic structural integrity of the bulk phases. A basic issue in formulating any molecular theory of macroscopic self-organisation of such complex molecular systems is the extent of detail to which the structure, the conformations and the interactions of the molecular units are to be described. In our previous works[51-53] we have introduced a general approach to coarse-grained modelling of complex molecular architectures together with the statistical mechanical theoretical framework for the exploration of possible order-disorder phase transitions. In this work, we extend our previous simple and tractable molecular theory to describe binary mixtures of anisotropic particles. The motivation for this work is twofold: On one hand there is a wealth of experimental knowledge on mesophase formation and demixing phenomena in anisometric particle mixtures from the fields of colloid[2-6,13-15], polymer[2-6] and liquid crystal[2-6,13-15,42] sciences which show interesting similarities. On the other hand there are numerous molecular theory[17,20-23,40,42-49] and simulation[19,22,24-28,38,40,42] studies, based on a variety of molecular models and statistical mechanics approaches, which successfully reproduce particular experimental observations, thus posing the challenge of seeking a unified yet simple theoretical description of the broad range of experimentally observed mesophase structures and multiphase equilibria. Here we address this challenge by applying a minimal modelling of the molecular shapes and primary molecular interactions in the context of a simplified statistical treatment wherein the molecular configurations are mapped on a cubic lattice. Considering the simplicity of the approach, an impressively successful qualitative reproduction of the complete phase behaviour in a rich variety of experimentally studied binary mixtures which combine rod-like, plate-like and spherical particles is obtained.

In the next section we present the molecular modeling and statistical mechanical framework of the calculations for mixtures of anisometric particles. In section 3 we present results for rod-sphere, plate-sphere and rod-plate binary mixtures; we also discuss the implications of shape anisometry, relative particle sizes and particle softness on the phase behavior of these mixtures and we compare our theoretical results with the experiment. The main findings are summarized and the conclusions of our theoretical study are drawn in Section 4.

## 2. Molecular modelling and statistical mechanics on a cubic lattice

Our study concerns binary mixtures in which one or both components are of anisometric molecular shape, namely rod-like or plate-like. Each component is assumed to be divided into a number of blocks, representing well defined submolecular segments, to which specific interactions are assigned. The total interaction potential between a pair of molecules is obtained as the sum of the interactions between all possible intermolecular pairs of blocks. To expedite the computations, a molecule is modeled as an array of cubic blocks that is allowed to move and rotate on a cubic three dimensional lattice. Thus, a rod-like molecule is represented by a linear string of a number $L_r$ of cubic blocks of type $r$ (rod). Similarly, a plate-like molecule is represented by a square two dimensional array of $L_p \times L_p$ cubic blocks of type $p$ (plate). The $r$ as well as the $p$ blocks are treated as impenetrable objects endowed with purely hard body interactions. A molecule of fully isometric shape, for brevity a spherical molecule, is represented in this model by a central cubic block $c$, which corresponds to the impenetrable core, surrounded by a layer of type $s$ blocks forming a cubical soft shell. In other words a spherical molecule is represented by a cube of $3\times3\times3$ blocks, one of them is a $c$ and the other 26 are $s$ blocks. The soft shell blocks $s$ are allowed to occupy the same lattice position with a block of type $r$ or $p$ and these simultaneous occupations are assumed to increase the potential energy of the system by a fixed amount $u$ per pair of overlapping $s-r$ or $s-p$ blocks. For simplicity, the overlapping of $s-s$ and $s-c$ block pairs is allowed without any change in the intermolecular potential energy. Thus, altogether the interaction potential $U_{a,b}$ for a given configuration of a pair of molecules $a,b$ (which can be any of the rod, plate or sphere type) receives an infinite contribution if any of the hard blocks (i.e. of type $r$ or $p$ or $c$) on the two molecules happen to occupy the same lattice position; these are the forbidden configurations. For the allowed configurations, $U_{a,b}$ receives a contribution $u$ for each intermolecular pair of $s$-$r$ or $s$-$p$ blocks that happen to coincide on a lattice site whilst all the other relative configurations of the two molecules have no effect on $U_{a,b}$. It should be noted that the inclusion of soft repulsive interactions between $s$-$r$ and $s$-$p$ pairs leads indirectly to effective attractions between molecules of the same kind in a rod-sphere or plate-sphere binary mixture.

The statistical mechanics is handled following the general approach of references 51-53 which we here extend to the case of binary mixtures. For the effectively rigid molecules considered in the present study, the extension is straight forward and leads to the following approximate expression for the Helmholtz-free energy:

$$\frac{F}{Nk_BT} = -\sum_{a=1}^{2} x_a \left\{ \ln \int d\varpi_a \zeta_a(\varpi_a) + \frac{1}{2}\sum_{b=1}^{2}(N_b - \delta_{a,b}) \times \int d\varpi_a \int d\varpi_b \rho_a(\varpi_a)\rho_b(\varpi_b)\exp(-U_{a,b}(\varpi_a,\varpi_b)/k_BT) \right\} \quad (1)$$

where, $T$ is the temperature, $k_B$ is the Boltzmann constant, $N = N_1 + N_2$ is the total number of particles, $x_a = N_a/N$ is the concentration and $\zeta_a(\varpi_a)$ is the variational weight function of the molecules of species $a$ (rod, plate or sphere, and similarly for the index $b$). The variable $\varpi_a = (R_a, \Omega_a)$ denotes collectively the position $R_a$ and orientation $\Omega_a$ of a molecule of species $a$. The function $\rho_a(\varpi_a)$ denotes the single molecule probability distribution of the such a molecule and is simply the normalized form of $\zeta_a(\varpi_a)$, i.e. $\rho_a(\varpi_a) = \zeta_a(\varpi_a) / \int d\varpi_a \zeta_a(\varpi_a)$. The variational weight functions $\zeta_a(\varpi_a)$ are determined self-consistently by functional minimization of the free energy eqn (1), leading to

$$\zeta_a(\varpi_a) = \exp\left[(N-1)\left(\sum_{b=1}^{2} x_b \frac{\langle \exp(-U_{a,b}(\varpi_a,\varpi_b)/k_B T)\rangle_b}{\langle \exp(-U_{a,b}(\varpi_a,\varpi_b)/k_B T)\rangle_o} - 1\right)\right] \quad (2)$$

with $\langle \exp(-U_{a,b}(\varpi_a,\varpi_b)/k_B T)\rangle_b = \int d\varpi_b \rho_b(\varpi_b) \exp(-U_{a,b}(\varpi_a,\varpi_b)/k_B T)$.

For given number density $N/V$, temperature $T$ and concentrations $x_a$, the self consistency eqn (2) are solved for functional forms of the weight functions $\zeta_a(\varpi_a)$ pertaining to the symmetries of the isotropic, nematic, orthogonal smectic and columnar fluid phases. For simplicity, uniaxial symmetry is considered for all the orientationally ordered phases. Due to the assumed form of the molecular interactions, the temperature enters in the self-consistency equations only through the dimensionless combination $u^* = u/k_B T$, which therefore is used as a scaled inverse temperature parameter. The same parameter can be regarded as a measure of the effective softness of the shell surrounding the hard core of the spherical molecules. For each set of solutions $\zeta_a(\varpi_a)$ we evaluate the free energy, the pressure and the chemical potentials at the respective density, concentration and temperature. The relative thermodynamic stability of the phases corresponding to the different solutions is determined by comparing the respective free energies. Demixing into coexisting phases I and II is identified through the simultaneous equality of the pressures, $P^I = P^{II}$, and the chemical potentials, $\mu_a^I = \mu_a^{II}$ ($a=1,2$), of each molecular species in the two phases.

## 3. Results and Discussion

The approach described in the previous section has been applied to three types of binary mixtures, namely rod-sphere, plate-sphere and rod-plate mixtures. The results are presented and discussed below for each binary mixture type separately.

**3.1. Rod-sphere mixtures**

Pressure-concentration phase diagrams have been calculated for rods of length $L_r$ forming binary mixtures with spheres consisting of a hard core block $c$ surrounded by a shell of soft blocks $s$ of effective softness $u^*$. Representative diagrams for combinations of lengths $L_r = 5$ and $L_r = 9$ and softness parameters $u^* = 0$ (no soft shell at all) and $u^* = 0.02$ (representing, for example, spheres with surface modification that renders them chemically incompatible to the rods) are shown in Fig. 1. Isotropic $(I)$, nematic $(N_r)$, smectic $(Sm)$ phases are found. The concentration is expressed in terms of the fraction $x_r = N_r/N$ of rod molecules in the mixture. The dimensionless parameter $P^* = P/P_{IN_r}$ is used for the pressure, with $P_{IN_r}$ denoting the value of the pressure at the isotropic to nematic $(I-N_r)$ phase transition of the pure rod system $(x_r = 1)$. It is apparent that the four diagrams of Fig. 1 have the same topology. At low pressures the systems are isotropic and fully miscible. On increasing the pressure stable nematic and smectic phases appear at high rod concentrations, separated by a miscibility gap from the isotropic mixture. The width of the miscibility gap, i.e. the range of compositions for which the system is biphasic, increases with increasing pressure. Analogous theoretical results in a mixture of rods and cubic particles have been obtained by Alben[20] in his pioneering study. A similar behavior has also been observed in Monte Carlo simulations of a mixture of hard Gaussian overlap particles[25]. There is a critical three-phase coexistence pressure, $P^*_{I-N_r-Sm}$, at which the system decomposes into coexisting isotropic, nematic and smectic phases with different compositions. The miscibility gap separates stable isotropic from stable nematic phases for pressures lower than $P^*_{I-N_r-Sm}$ and separates stable isotropic from stable smectic phases at higher pressures. At high rod concentrations a nematic to smectic $(N_r - Sm)$ phase transition is obtained. A qualitatively similar phase behavior to the one shown in Fig. 1 has been found in reference 30 for mixtures of hard spherocylinders and spheres.

It is clear from the phase diagrams of the binary mixtures with $u^* = 0$ that the range of compositions for which the systems self organize into stable liquid crystalline phases increases on increasing the rod length. Furthermore the nematic phase for the long rod ($L_r = 9$) system is stable for pressures up to nine times higher than the pressure $P_{IN_r}$. In contrast, the short rod system ($L_r = 5$) undergoes a transition to the smectic phase at about four times $P_{IN_r}$. It is also worth noting in the phase diagrams Fig. 1b, 1c and 1d that there is a range of composition for which the mixture exhibits a pressure-induced reentrant demixing behavior. This type of reentrant behavior has also been obtained theoretically[29] and in Monte Carlo simulations[25] for rod-sphere mixtures.

The phase diagrams with $u^* = 0.02$ show a wider miscibility gap; this is an expected consequence of increasing the incompatibility between the two components of the mixture through the soft shell repulsions. For the same reason the stable ordered phases are shifted towards high rod concentration region of the phase diagram and the pressure induced demixing re-entrance behavior is enhanced. Particularly noteworthy is the crossover of the slope of the nematic-smectic transition line from clearly negative for $u^* = 0$ to clearly positive for $u^* = 0.02$. To our knowledge, such switching of the slope has not been observed or predicted in any of the previous studies. A positive (negative) slope of the $N_r - Sm$ line indicates that the nematic phase is destabilized relative to the smectic phase on increasing (decreasing) the concentration of the spherical molecules in the mixture. On the other hand, it has been found[22], both in molecular theory and in computer simulations of mixtures of spheres and parallel spherocylinders, that the addition of spherical particles stabilizes the layered phases. This however is not in contradiction with the switching of the slope shown in Fig. 1, since the calculations in reference 22 refer to constant total volume-fraction conditions. Under such conditions, our calculations indicate a stabilization of the layered phase on increasing the spherical component concentration, in agreement with previous predictions[21-23]. This is apparent from the positive slope of all the graphs of Fig. 2, where the calculated packing fraction $\eta_{N_r-Sm} = (N/V)(x_r v_r + x_c v_c)_{N_r-Sm}$ at which the $N_r - Sm$ phase transition occurs is plotted as a function of the composition of the rods for the system with $u^* = 0$ and various rod lengths. In the expression of the packing fraction, $x_r, x_c$ and $v_r, v_c$ denote respectively the concentrations and molecular volumes of the rod and sphere components of the mixture.

The calculated nematic order parameter, $S = \langle 3(\mathbf{n} \cdot \mathbf{e})^2 - 1 \rangle / 2$, with $\mathbf{n}$ and $\mathbf{e}$ denoting the nematic director and the molecular orientation respectively, indicate that the nematic phase starts out in the phase sequence with $S$ in the range $0.80-0.86$, with the higher ordering corresponding to higher sphere compositions. The transition from the nematic to the smectic phase take place at extremely high values of $S$ (>0.99) indicating that rods are practically parallel near the $N_r - Sm$ phase transition.

Lastly, from the positional distribution of the rods and spheres, it follows that the smectic phase consists of alternating rod-rich and sphere-rich layers. Such a smectic phase, termed as lamellar, has been observed[2-4] experimentally in suspensions of colloidal rods and spheres. The calculated probabilities for spheres in the rod-rich region are in general small but not negligible. It is also found that the layer spacing of the smectic phase decreases with increasing pressure and rod concentration $x_r$. For small sphere concentrations the layer spacing does not change dramatically. Experimental results on polystyrene spheres and fd virus ("rods" of length $\sim 900 nm$) do not show swelling of the layers as a function of sphere concentration[2].

### 3.2. Plate-sphere mixtures

Pressure-concentration phase diagrams have been calculated for binary mixtures of plates of size $L_p \times L_p$ with spheres consisting of a hard core block $c$ surrounded by a shell of soft blocks $s$ of effective softness $u^*$. Isotropic $(I)$, plate nematic $(N_p)$, and columnar $(Col)$ phases are found.

For purely hard body interactions ($u^* = 0.00$) the phase diagrams have the topology shown in Fig. 3. These are similar to the diagrams of Fig. 1, with the $N_p$ and *Col* in place of the $N_r$ and *Sm* phases, respectively. The pressure is again expressed in units of the $I - N_p$ transition pressure $\left(P_{IN_p}\right)$ of the pure plate system. Increasing the shape anisometry of the plates, by increasing $L_p$, has analogous consequences with increasing the length $L_r$ in the rod-sphere mixtures, namely (i) expansion of the stability range of the nematic phase relative to the isotropic and columnar phase (ii) shrinking of the coexistence regions of the isotropic phase with either of the ordered phases and (iii) shift of the miscibility gap towards higher concentrations of the sphere component. The negative slope of the nematic to columnar phase transition line in the pressure-composition diagrams indicates that, under constant pressure, the columnar phase is destabilized in favor of the nematic on adding spherical particles. In analogy with the hard rod-sphere mixtures, this trend is reversed if constant volume conditions are applied. In this case the addition of hard spheres enhances the stability of the columnar phase but at the same time higher pressures are required to maintain the mixture volume constant.

To analyze the molecular organization in the columnar phase we have calculated the orientationally averaged two dimensional density profiles of the plates and spheres on a plane perpendicular to the columns. Representative results are shown in Fig. 4. The density profiles clearly show that the plate molecules stack up to form columns and that these columns form a rectangular lattice with the corridors in between the columns populated by spheres.

Endowing the spherical molecules with an outer shell that softly repels the plate molecules, by taking $u^* > 0$, we obtain the phase diagrams of Fig. 5. For small $u^*$ the topology of the phase diagram shows only quantitative differences with the diagram of Fig. 3a, particularly in the breadth of the miscibility gap. For large $u^*$ (Fig. 5b) the topology of the phase diagram changes and the system exhibits two different isotropic phases ($I_1$, $I_2$) that coexist at low pressures over a wide range of compositions. The two isotropic phases differ in the relative concentrations of plates and spheres. At higher pressures the system becomes practically immiscible and for pressures above $P^* > 3.9$ the system appears to decompose into plates that form a columnar phase and spheres that form an isotropic fluid.

The phase diagrams in Fig. 5 are relevant to the phase sequences observed experimentally[13] and studied theoretically[37-40] for mixtures of disk-like colloids with non-adsorbing polymers[13] where, aside from the widening of the demixing gap, an additional isotropic phase was observed. Apparently, the repulsive interactions among unlike particles introduce an effective attraction, known as depletion attraction[54,55], among like particles.

### 3.3. Rod-plate mixtures

Experimental results on mixtures of rod-like and plate-like colloidal particles have revealed a rich phase behavior[14,15]. In these mixtures, the mutual excluded volume of a pair of plates is much larger than that of a pair of rods. In addition to distinct nematic phases and respective coexistence regions, a plate rich columnar phase appears. The phase behavior of a model of hard boehmite rods and gibbsite platelets was explored theoretically[15,41] in the context of an Onsager type approach. These theoretical results describe correctly the experimentally observed phase behavior at low particle densities. However, the possibility of formation of positionally ordered liquid crystal phases (smectic and columnar) at higher densities is not considered in these calculations. Here we address this point within our molecular modeling.

To convey the excluded volume differences of the experimental systems we have chosen rods of length $L_r = 13$ and square plates of side length $L_p = 9$. For these particle dimensions, the excluded volume for a pair of plates, $q_{pp}$, is nine times the excluded volume for a pair of rods, i.e. $q_{pp} = 9q_{rr}$, and the excluded volume for a rod-plate pair is $q_{rp} = 4q_{rr}$. Accordingly, a strongly asymmetric phase diagrams with respect to the relative composition variable is to be expected. The calculated pressure-concentration phase diagram for this system is presented in Fig 6a. The pressure is expressed in units of the $I - N_p$ transition pressure $\left(P_{IN_p}\right)$ of the pure plate system. Four different phases - one isotropic phase, two uniaxial nematic phases (a rod-rich $N_r$ and a plate-rich $N_p$) and a plate rich columnar phase $(Col)$ - and respective phase coexistence lines are found. Phase coexistence is indicated by horizontal tie lines. In the columnar phase the plates and rods assemble into columns. At low plate concentrations alternating thick and thin bands, containing plates and rods respectively, are formed in the x-y plane. At higher plate concentrations, the plates pack into columns which form a two-dimensional tetragonal lattice whilst the rods form crossing corridors in the space left between the plate columns. In both cases the long axis of the rods is preferentially oriented parallel to the x-y plane, i.e. perpendicular to the columnar axes.

A particularly interesting feature of the phase diagram in Fig. 6a is the presence of two triple points: the one is at $(x_p = 0.032; P^* = 7.0)$, where the isotropic phase coexists with the two nematic phases $(I - N_r - N_p)$; the other point is at $(x_p = 0.47; P^* = 13.2)$ where the columnar phase coexists with the two nematic phases $(N_r - N_p - Col)$. Compressing the mixture at low plate concentrations, say $x_p \approx 0.1$, a notable sequence of phase transitions takes place: from $I$ to $I + N_p$, $I + N_r + N_p$ to $N_r + N_p$ to $N_r + N_p + Col$ and finally to high-pressure $N_r + Col$ demixing.

Fig. 6b presents the calculated phase diagram in terms rod volume fraction $(\eta_r)$ vs plate volume fraction $(\eta_p)$ variables at phase coexistence. This representation is useful because it allows a straightforward comparison with the experimental results[14] (see Fig. 6c). The triphasic equilibrium corresponds to the two distinct triangular regions of the phase diagram of Fig. 6b. Our theoretical results are in qualitative agreement with the experimental observations in the low density regime. In particular, the theoretically predicted phase behavior describes very well the coexistence between the isotropic and the uniaxial rod-rich and plate-rich nematic phases. The $I - N_r - N_p$ triphasic region, which clearly appears in the experimental phase diagram, is also accounted for by the calculation. On the other hand, the experimental observations show, beyond the triphasic region, two high density liquid crystalline phases, a plate rich columnar and a rod rich phase X. Thus, the theoretically predicted $N_r - N_p$ demixing appears to be preempted by multiphase equilibria of more than three phases. This deviation can perhaps be attributed to polydispersity in the sizes and shapes of the experimental colloidal particles. This is in accord with the Gibbs phase rule whereby in a truly binary mixture no more than three phases can coexist. Aside from this deviation, the theoretical phase diagram reproduces correctly the coexistence of a rod–rich nematic with a plate-rich columnar phase at high densities and also the $N_p - Col$ coexistence at low $\eta_r$ and high $\eta_p$ as well as the triphasic $N_r - N_p - Col$ region appearing the experimental phase diagram.

The structure of the phase diagrams show considerable sensitivity to the relative sizes of the rod and plate components. To explore this we have considered different combinations of sizes. Fig. 7a presents the calculated pressure- concentration phase diagram for a binary mixture of size $L_p = 5$ plates with size $L_r = 7$ rods. The low pressure region corresponds to the isotropic phase over the entire concentration range. A triphasic $(I - N_r - N_p)$ equilibrium is obtained at the triple point $\left(x_p = 0.056, P^* = 2.8\right)$. Narrow $I - N_r$ and $I - N_p$ coexistence regions appear on either side of the triple point as the pressure is increased or decreased, respectively. A reentrant transition is observed at plate concentrations between 0,07 and 0,11. In that range of concentrations, increasing the pressure produces the following sequence of phases: from $I$ to $I + N_p$ to $N_p$ to $N_r + N_p$ to $N_r$ then again to $N_r + N_p$ to $N_r + N_p + Col$ and finally high pressure $N_r + Col$ decomposition. At intermediate pressures the diagram shows a demixing region into two uniaxial nematic phases, a rod-rich $(N_r)$ in coexistence with a plate-rich $(N_p)$ nematic. Furthermore, a biaxial nematic mixture of rods and plates is also found in that region. This phase, however, is thermodynamically unstable to demixing into the two uniaxial nematic phases and is therefore not shown on the phase diagram. This point has been investigated extensively[16] and will not be further considered here. A triphasic $(Col - N_r - N_p)$ equilibrium is obtained at the second triple point $(x_p = 0.44, P^* = 5.6)$ of the phase diagram of Fig7a.

Fig. 7b presents the phase diagram for binary rod-plate mixtures only with more elongated rods, namely $L_r = 11$, and plates of $L_p = 5$ as in Fig 7a. This combination of molecular sizes yields relative values of the molecular-pair excluded volumes $q_{rp} = 1.61 q_{rr}$ and $q_{pp} = 2.02 q_{rr}$. The topology of the two phase diagrams in Fig. 7a and 7b is similar, except that the triple points are found at higher concentrations and no phase reentrance transition is obtained for the phase diagram in Fig 7b.

## 4. Conclusions

In this work we have introduced a theoretical model of shape anisometric particles embedded in a cubic lattice and have used it to describe binary mixtures combining rod-like, plate-like and spherical particles.

For the mixtures of rod-like with spherical particles we have found thermodynamically stable isotropic, nematic and lamellar phases as well as triphasic coexistence between these phases and regions of biphasic coexistence of all possible combinations of the three phases. The lamellae form a succession of rod-rich and sphere-rich layers. We have also found that the presence of soft shell repulsions between the rods and the spheres tends to destabilize the lamellar phase relative to the nematic.

For mixtures of plate-like with spherical particles we have found isotropic, nematic and columnar phases. In the latter, the columns are formed by plate-like particles and the spheres occupy the inter-columnar regions. Triphasic and biphasic equilibria between the three phases are found. The presence of soft shell repulsions between the rods and the spheres tends to produce two different isotropic phases in coexistence over a relatively wide range of compositions.

In both of the above types of binary mixtures, increasing the anisometry of the rods or plates leads to the enhancement of the stability of the nematic phase, relative to both the isotropic and the positionally ordered (lamellar or columnar) phases, together with shrinking of the concentration range of the coexistence of the isotropic phase with the ordered phases and the displacement of the miscibility gap towards higher sphere concentrations.

The phase polymorphism is further enriched in mixtures of plate-like with rod-like particles. The thermodynamically stable phases include an isotropic, a columnar and two types of uniaxial nematic phases. Triphasic and biphasic coexistence is obtained for several combinations of these four phases. In the columnar phase the columns are formed by the plates while the rods are found to occupy the space between the columns and to be preferentially aligned perpendicular to the axes of the columns. The phase diagrams are found to be fairly sensitive to the relative sizes of the rod and plate particles.

Despite its extreme simplicity regarding the shapes and interactions of the particle as well as their allowable spatial configurations, the model reproduces remarkably well the qualitative experimental trends in all the binary mixture combinations studied and is also in agreement with a broad variety of theoretical and computational results obtained for specific types of binary mixture. This suggests that, behind the apparent richness and diversity, mesophase structure and thermodynamic stability in all these binary mixtures might be controlled by very basic microscopic attributes such as particle shape-anisometry and relative size, which determine the hard body packing properties, perhaps combined in some cases with possible soft shell interactions which could also confer a more sensitive dependence on temperature.

**FIGURE CAPTIONS**

**Figure 1**. Calculated pressure ($P^*$)-concentration ($x_r$) phase diagrams for rod-sphere binary mixtures. Four combinations of rod length $L_r$ and softness parameter $u^*$ of the spherical shell are shown: (a) $L_r = 5$, $u^* = 0.00$; (b) $L_r = 5$, $u^* = 0.02$; (c) $L_r = 9$, $u^* = 0.00$; $u^* = 0$, (d) $L_r = 9$, $u^* = 0.02$. Cartoons of the molecular self organization are shown for the marked points on the phase diagrams.

**Figure 2.** Calculated packing fraction $\eta_{N_r-Sm}$ at the $N_r - Sm$ phase transition as a function of the rod concentration $x_r$ for a binary system of rods and spheres interacting via purely hard body repulsions ($u^* = 0$). Results are shown for different lengths $L_r$ of the rod component.

**Figure 3.** Calculated pressure-concentration phase diagrams for plate-sphere binary mixtures interacting via purely hard-body repulsions ($u^* = 0$). Results are shown for two sizes of the square plates: (a) $L_p = 5$ and (b) $L_p = 7$. Cartoons of the molecular self organization are shown for the marked points on the first phase diagram.

**Figure 4.** Density profiles of (a) the plate component and (b) the sphere component for the binary mixture of fig. 3a (i.e. $u^* = 0$ and $L_p = 5$) at plate concentration $x_p = 0.43$ and pressure $P^* = 9$.

**Figure 5**. Calculated pressure-concentration phase diagrams for plate-sphere binary mixtures with hard core and soft shell interactions of softness $u^* = 0.01$ (a) and $u^* = 0.2$ (b). The size of the square plates is $L_p = 5$ in both cases.

**Figure 6.** (a) Calculated pressure vs plate concentration phase diagram for plate-rod binary mixtures of molecular sizes $L_p = 9$ and $L_r = 13$. Cartoons of the molecular self organization are shown for the marked points on the phase diagram. (b) Calculated phase diagram using as variables the packing fraction of rods ($\eta_r$) vs packing fraction of plates ($\eta_p$) as applicable to the phase boundaries of (a). (c)

Experimental phase diagram for mixtures of colloidal boehmite rods (aspect ratio $L_r / D_r \sim 10$) and gibbsite platelets (aspect ratio $D_p / L_p \sim 15$), redrawn after reference 14.

**Figure 7.** Calculated pressure-concentration phase diagrams for rod-plate binary mixtures in which the plates are squares of side $L_p = 5$ and the rods have length either (a) $L_r = 7$ or (b) $L_r = 11$.

2323# FIGURES

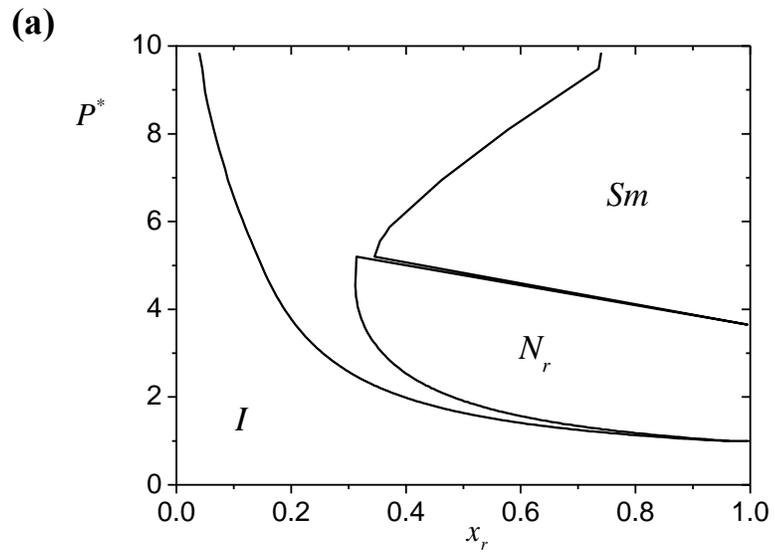

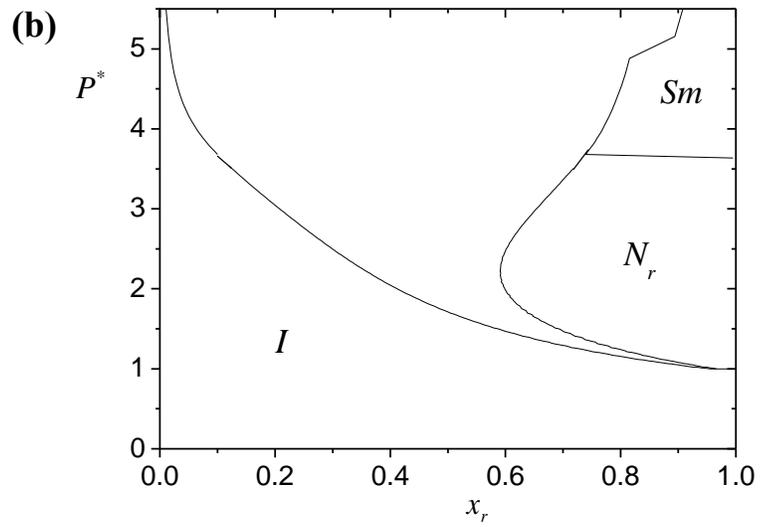

**Figure 1**

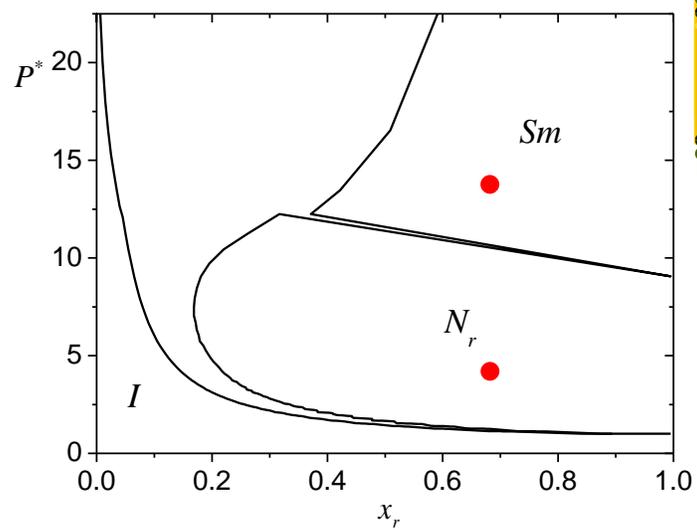

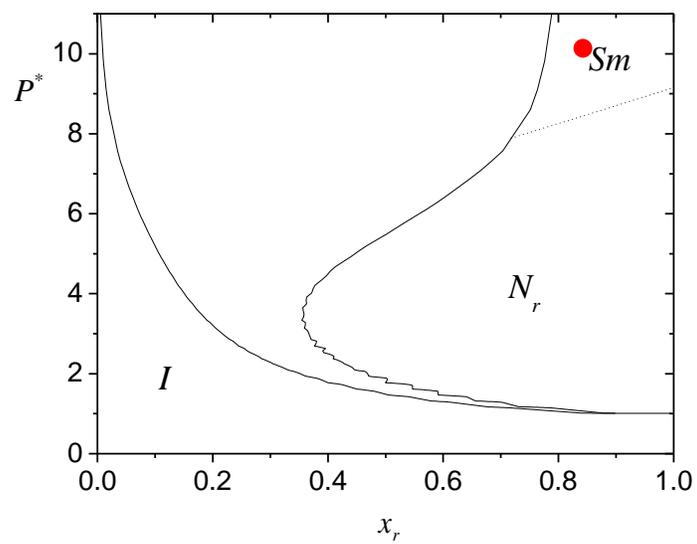

**Figure 1**

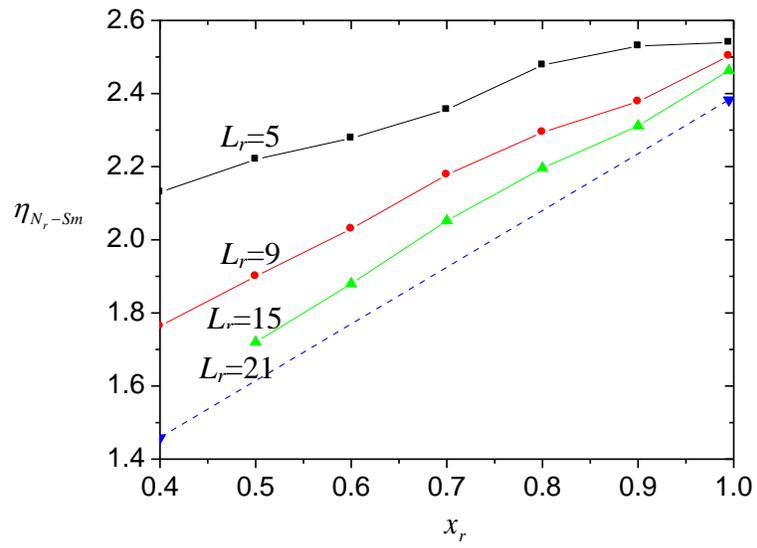

**Figure 2**

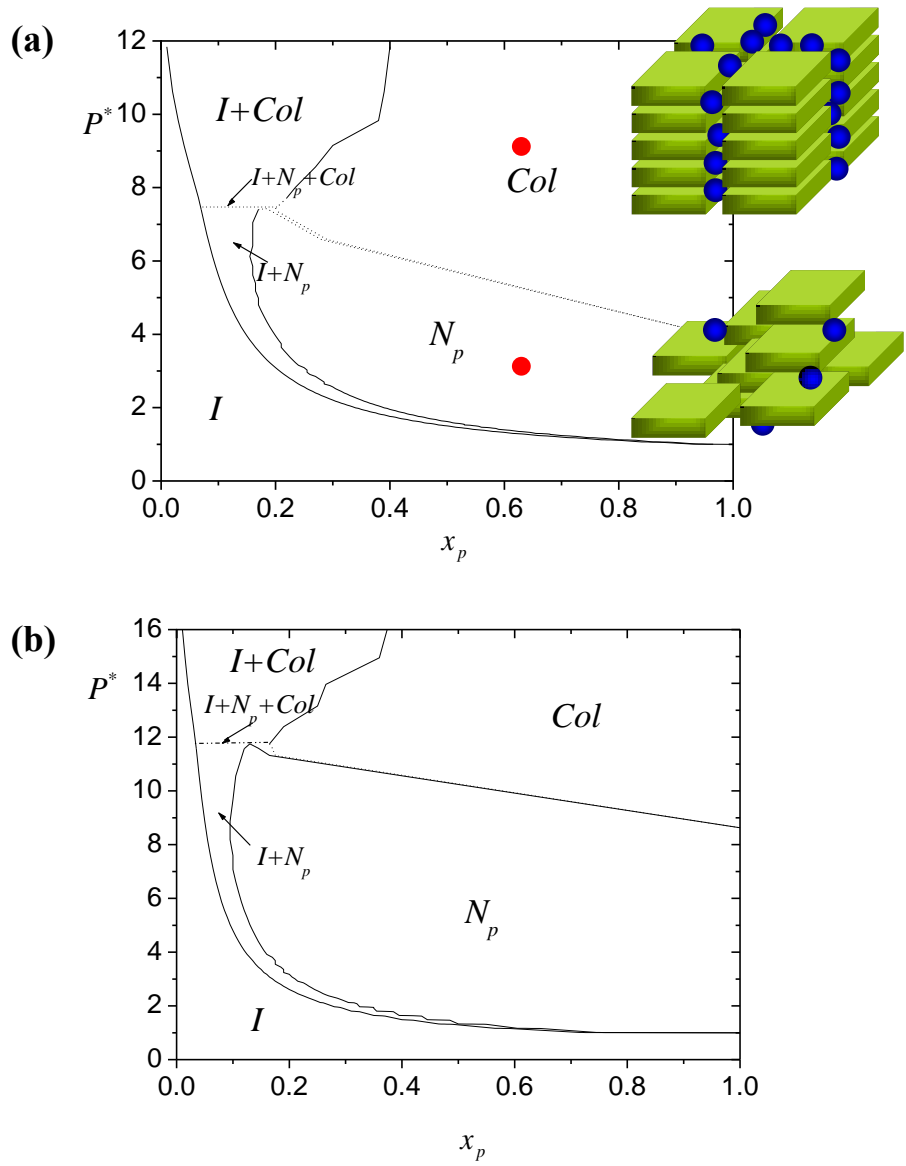

**Figure 3**

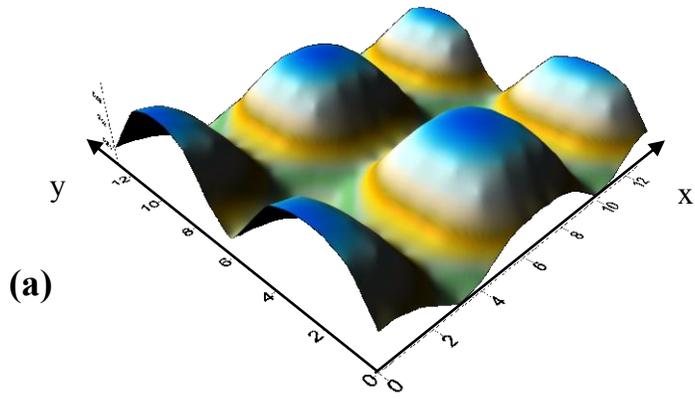

(a)

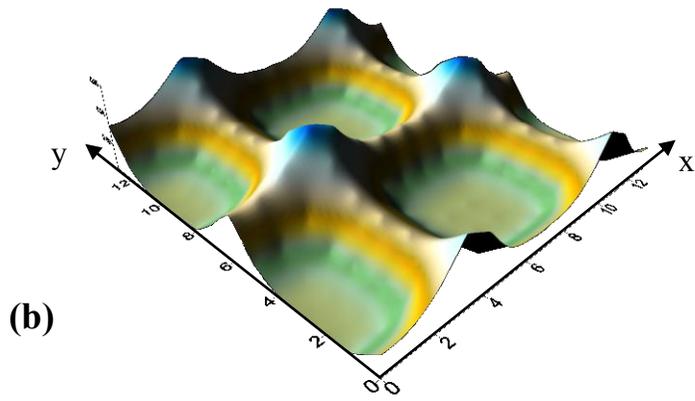

(b)

**Figure 4**

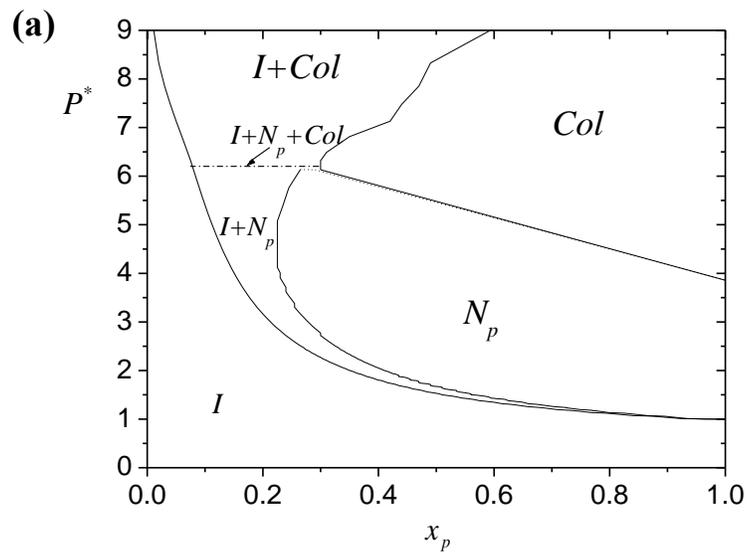

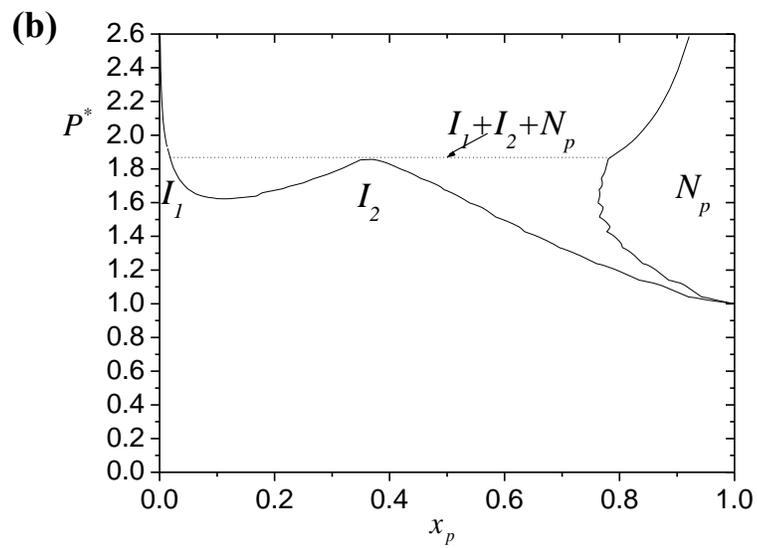

**Figure 5**

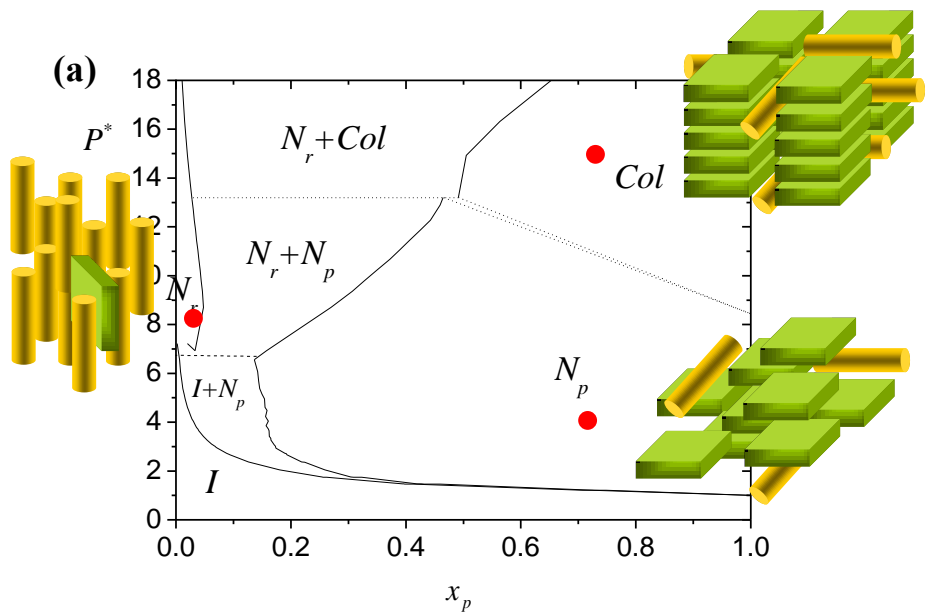

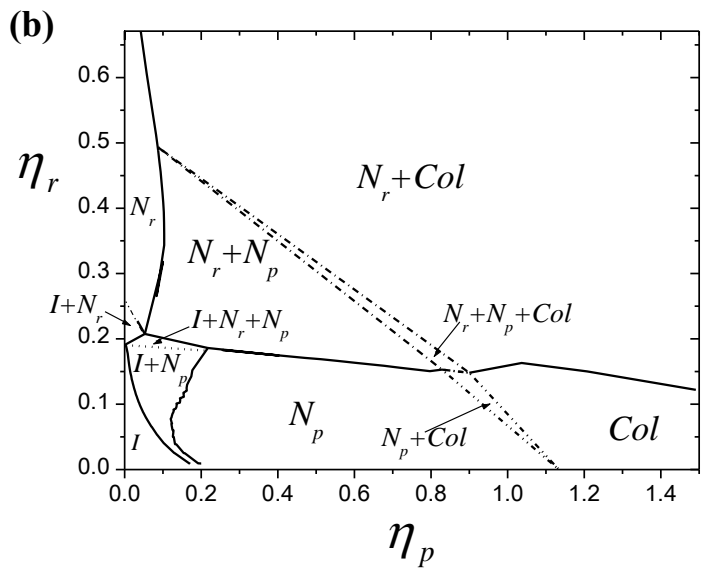

**Figure 6**

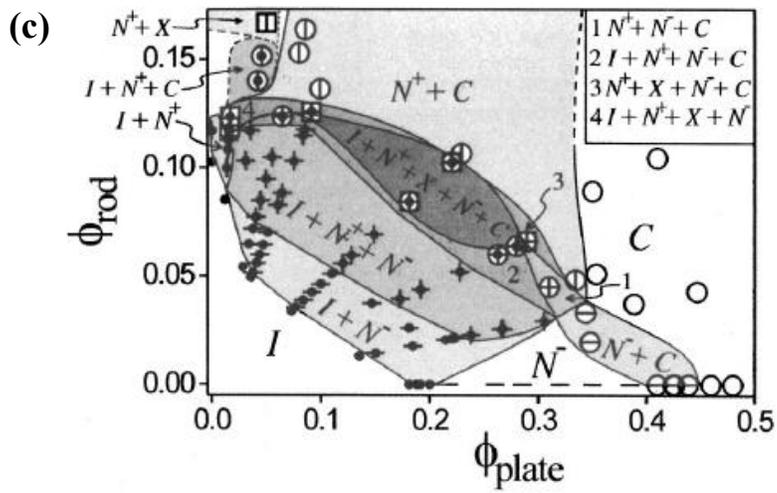

**Figure 6**

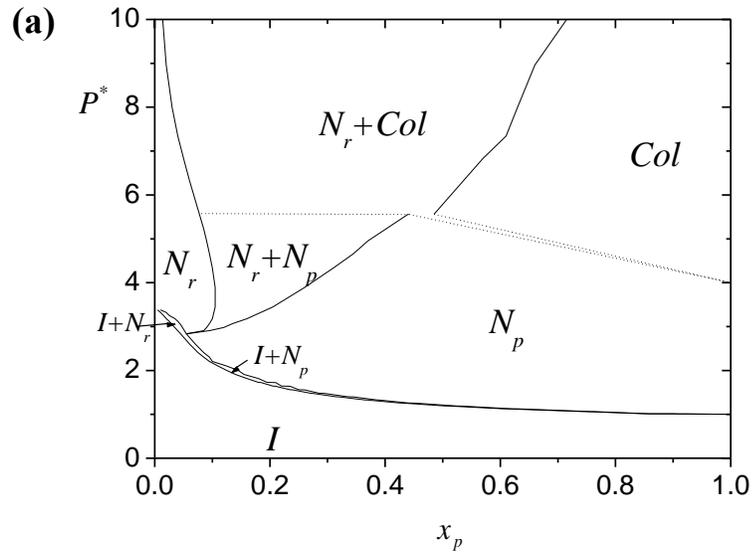

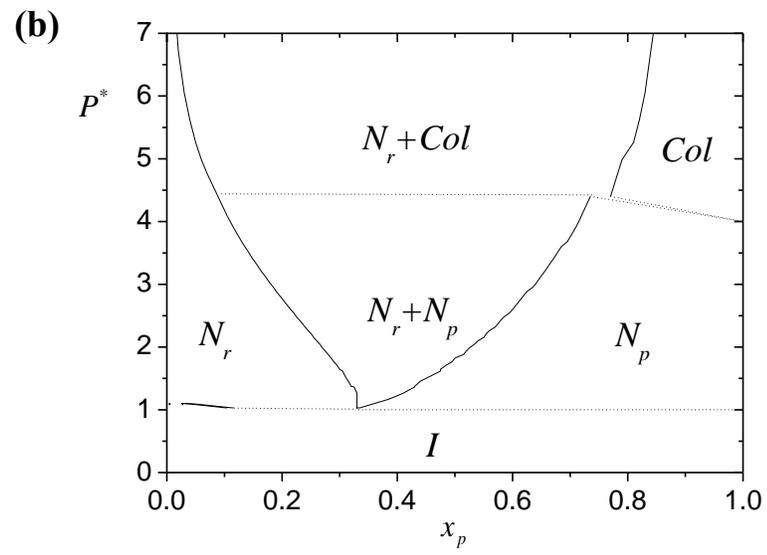

**Figure 7**